\documentclass[aps,pra,twocolumn,groupedaddress,noeprint,nodoi]{revtex4-2}
\usepackage{amsmath}
\usepackage{physics}
\usepackage{graphicx}  %for arXiv
\newcommand{\ii}{{\mathrm{i}}}
\newcommand{\ee}{{\mathrm{e}}}
\begin{document}
\title{BIC lasing from first principles}
\author{Tetsuyuki Ochiai}
\affiliation{Research Center for Electronic and Optical Materials,  National Institute for Materials Science (NIMS), Tsukuba 305-0044, JAPAN}
\date{\today}

\begin{abstract}
One of the main applications of the bound state in the continuum (BIC) is the low-threshold lasing. Ideally, the infinite quality factor of the BIC results in the zero-threshold and zero-linewidth lasing. However, various effects disturb the infinite quality factor. Instead, we have a dense distribution of eigenmodes around a target frequency of ideal systems, giving rise to a complex mode selection under finite pumping. We here study how the lasing modes are selected in finite photonic crystals composed of dielectric spheres with gain media. We employ a semi-classical approach, the so-called steady-state ab-initio laser theory combined with the multiple-scattering method. 
The theory predicts that the BIC certainly results in the low-threshold lasing in the finite systems. Moreover, the localfield enhancement of the BIC stabilizes the single-mode lasing far beyond the higher lasing threshold in the noninteracting limit.  
\end{abstract}

%\keywords{}
\maketitle

\section{Introduction}

The optical bound state in the continuum (BIC) is a resonant mode with infinite quality factor ($Q$) in a photonic structure \cite{Miyazaki1998a,Paddon:Y::61:p2090-2101:2000,Ochiai:S::63:p125107:2001,Fan:J::65:p235112:2002,Marinica2008,Plotnik2011,Hsu2013b}.   The radiation field is strongly confined inside the structure, and acts as an ideal cavity.  The strong confinement enables us to study various enhanced linear and nonlinear optical effects \cite{Kang}.

Among them, we here focus on lasing via BICs \cite{Ryu2002,Kodigala2017,Ha2018,Hwang2021a}.   Not only the zero threshold and zero quantum linewidth \cite{Schawlow1958} due to the infinite $Q$, but also the strong nonlinearity make the BIC lasing distinctive from conventional lasing, because the optical nonlinear interaction in the gain saturation is essential in the lasing.

In practice, however, many factors hide the BIC and infinite $Q$.  
Finite size effects, energy dissipation, and structural disorder are representative ones. Since most BICs are found in specific Bloch modes in photonic crystal membranes, 
finite system sizes blur the photonic bands with definite Bloch waves. Instead, we have a dense distribution of photonic eigenmodes in the frequency regions of the photonic bands. 
It is nontrivial how the BICs are formed from the resonances in the finite systems \cite{Hoang2024a}.

For instance, the BICs are typically found at the $\Gamma$ point, which is usually at a band edge. 
In finite systems, the band edges are formed by 
a one-sided accumulation frequency of the resonance modes densely distributed toward it. 
The $Q$ factors differ mode by mode, becoming very high toward the band-edge frequency. Therefore, 
even if the highest $Q$ mode exhibits lasing, we may expect multimode behavior under a slight increase in the pumping. 
Thus, the ultra-low-threshold BIC lasing may exhibit a complicated transition from single to multimode lasing.

In this paper, we investigate theoretically how lasing occurs in finite systems that can hold BICs in the infinite system-size limit. A similar issue was investigated in terms of infinite periodic systems \cite{Benzaouia2020b}. 
We found that a single-mode lasing due to the symmetry-protected BIC is stabilized far beyond the multimodal threshold in its linear analysis. The lasing mode has a node by the symmetry-protected BIC.

To this end, we employ the so-called steady-state ab initio laser theory (SALT) \cite{Ge2010a} combined with the multiple scattering method \cite{Xu1995,DeAbajo1999}.
There are various approaches to lasing in photonic nanostructures, including the tight-binding model \cite{Harari2018}, coupled-wave theory combined with Langevin noise \cite{Inoue2019}, and the Maxwell-Bloch equation \cite{Oskooi2010}. 
Most of them are time-domain simulations. 
If we focus on the BIC, the frequency-domain approach better treats various BIC properties.

The SALT is a semi-classical laser theory based on the frequency-domain Maxwell-Bloch equation under the assumption of the steady-state population inversion. It can treat complex photonic environments in a first-principles manner. The multiple-scattering method is optimized for photonic systems made of spherical or cylindrical scatterers. Therefore, the lasing in the nanophotonic systems made of dielectric spheres or cylinders can be well-described with this approach. Moreover, such systems can have BICs \cite{Bulgakov2014,Yuan2017a,Bulgakov2017a,Bulgakov2018a,Kostyukov2022,Ochiai2024} in a controllable manner.

This paper is organized as follows. 
In Sec. II, we briefly summarize the SALT in a fully vectorial formulation. Section III is devoted to formulating the nanophotonic systems made of spherical particles as a lasing platform.  In Sec. IV, we consider the BIC lasing in a finite one-dimensional (1D) lattice of spheres. In Sec. V, 
a finite two-dimensional (2D) square lattice is investigated as the platform of the BIC lasing.  
Finally, in Sec. VI, summary and discussions are given.

\section{Steady-state ab initio laser theory with vectorial formulation}

We briefly summarize the SALT with the vectorial formulation. 
The vectorial formulation is necessary for spherical particle systems presented in the following sections.

The SALT equation is given by 
\begin{align}
&\nabla \times\nabla\times \vb*{\psi}^\mu(\vb*{x}) \nonumber\\
&\hskip5pt =\frac{\omega_\mu^2}{c^2} \left(\epsilon(\vb*{x})+\theta(\vb*{x})\frac{d_0\gamma(\omega_\mu)}{1+\sum_\nu | \gamma (\omega_\nu)\vb*{\psi}^\nu(\vb*{x})|^2 }\right)\vb*{\psi}^\mu(\vb*{x}),\\
& \gamma(\omega)=\frac{\gamma_\perp}{\omega-\omega_a+\ii\gamma_\perp},
\end{align}
where $\epsilon(\vb*{x})$ is the dielectric function of the system without the gain medium for lasing, $\theta(\vb*{x})$ is the pumping function of the gain medium that is 1 in the pumping region and 0 otherwise, $d_0=D_0|\vb*{d}_{21}|^2/(\hbar\gamma_\perp)$ is the normalized pumping rate, $D_0$ is the equilibrium population inversion ($n_2-n_1$ being $n_2$ and $n_1$ the densities of upper and lower levels, respectively, in the gain medium) under vanishing external radiation, $\vb*{d}_{21}$ is the dipole matrix element between the upper and lower levels the gain medium, $\omega_a$ is the level splitting between the two levels, and $\gamma_\perp$ is the transverse relaxation rate.    
The electric field of the lasing mode is given by 
\begin{align}
\vb*{E}(\vb*{x},t)=\Re\left[ \sum_{\mu=1}^{N_\mathrm{las}} \vb*{\psi}^\mu(\vb*{x})\ee^{-\ii\omega_\mu t} \right] 
\end{align}
in the SALT unit ($\frac{\hbar\sqrt{\gamma_\|\gamma_\perp}}{2|\vb*{d}_{21}|}$)
being $\gamma_\|$ the longitudinal relaxation rate of the gain medium.   
Here, $\mu$ represents each lasing mode,  $\omega_\mu$ is the real-valued lasing frequency of mode $\mu$, and $N_\mathrm{las}$ is the number of lasing modes. The SALT equation self-consistently determines them.

The SALT is a semi-classical laser theory neglecting the randomness of  the spontaneous emission, so that the linewidth of the lasing modes is supposed to vanish. 
Although an extension of the SALT can explain the linewidth \cite{Pick2015}, we here focus on possible effects of the BICs in the above quantities.

The SALT also assumes well-separated (and nondegenerate) lasing frequencies in multimode regions in comparison to $\gamma_\|$  \cite{Burkhardt2015,Liu2017a}. This  assumption becomes a crucial limitation in our systems presented in the following sections, because 
dense or degenerate lasing modes are available.  We thus mainly consider  nondegenerate single-mode regions.

To solve the SALT equation, we expand the lasing modes by the threshold constant flux (TCF) states defined by 
\begin{align}
	\nabla \times\nabla\times \vb*{E}_\eta(\vb*{x};\omega)=\frac{\omega^2}{c^2} \left(\epsilon(\vb*{x})+\theta(\vb*{x})\eta \right)\vb*{E}_\eta (\vb*{x};\omega),
\end{align}
where $\omega$ is the frequency of interest and $\eta$ is an eigenvalue to be determined.  The outgoing boundary condition is imposed. Namely, at far-field, the TCF states are expressed as 
\begin{align}
	\vb*{E}_\eta (\vb*{x};\omega)\simeq \frac{1}{r}\ee^{\ii q_br}\vb*{f}_\eta(\Omega;\omega), \quad q_b=\frac{\omega}{c}\sqrt{\epsilon_b}, \label{Eq:ff}
	\end{align}
being $\epsilon_b$ the dielectric constant of the background medium and $(r,\Omega(\theta,\phi))$ the polar coordinate. 
The complex eigenvalues $\eta$ are frequency-dependent and discrete in the lower half plane: $\Im[\eta]<0$.  

The TCF states are shown to be orthogonal to each other. 
We have 
\begin{align}
&\frac{\omega^2}{c^2}(\xi-\eta)\int \dd\vb*{x}\theta(\vb*{x})
	\vb*{E}_\xi (\vb*{x};\omega)\cdot 	\vb*{E}_\eta (\vb*{x};\omega)\nonumber\\
&=\ii\omega \int \dd\vb*{S}\cdot (	\vb*{E}_\xi (\vb*{x};\omega)\times \vb*{B}_\eta (\vb*{x};\omega) + \vb*{B}_\xi (\vb*{x};\omega)\times \vb*{E}_\eta (\vb*{x};\omega) ). 
\end{align}
The surface integral in the right-hand side vanishes at $r=\infty$ under Eq. \eqref{Eq:ff}. 
Thus, we can ortho-normalize the TCF states as  
\begin{align}
	\int \dd \vb*{x} \theta(\vb*{x})\vb*{E}_\xi(\vb*{x};\omega)\cdot\vb*{E}_\eta(\vb*{x};\omega)=\delta_{\xi,\eta}. \label{Eq:norm}
\end{align}

This normalization has a striking contrast to those in the quasi-normal modes (QNMs), which are often employed to study optical resonances \cite{Sauvan2013a}. The QNMs have complex eigenfrequencies while the far-field pattern is still expressed as Eq. \eqref{Eq:ff}. Since their eigenfrequencies are found in the lower half plane ($\Im[\omega]<0$), they grow exponentially at far fields. However, the QNMs can be orthonormalized using the magnetic field. 
We also note that the magnetic part of the inner product of the TCF states are not orthogonal.

Since the equation for the TCF states is the frequency-domain Maxwell equation with a slight modification, it can be solved with conventional frequency-domain Maxwell solvers.  In the following sections, we employ the multiple scattering method based on the vector spherical waves.

Expanding the lasing mode profile by the TCF states
\begin{align}
	&\vb*{\psi}^\mu(\vb*{x})=\sum_n c_n^\mu \vb*{E}_{\eta_n}(\vb*{x};\omega_\mu), \label{Eq:expandTCF}
\end{align}
the SALT equation becomes a nonlinear eigenvalue equation as  
\begin{align}
&c_n^\mu=\sum_{n'}d_0T_{nn'}^\mu(\{\omega_\nu,c_m^\nu\})c_{n'}^\mu, \label{Eq:NLE}\\
&T_{nn'}^\mu(\{\omega_\nu,c_m^\nu\})\nonumber\\
&\hskip5pt =\frac{\gamma(\omega_\mu)}{\eta_n(\omega_\mu)}\int \dd\vb*{x}\theta(\vb*{x}) \frac{ \vb*{E}_{\eta_n}(\vb*{x};\omega_\mu)\cdot\vb*{E}_{\eta_{n'}}(\vb*{x};\omega_\mu)}{1+\sum_\nu|\gamma(\omega_\nu)\sum_{m}c_m^\nu \vb*{E}_{\eta_m}(\vb*{x};\omega_\nu)|^2} 
\end{align}
This equation determines the lasing frequencies $\omega_\mu$ and coefficients $c_n^\mu$, self-consistently.  We employ the iterative method described in Ref. \onlinecite{Ge2010a} to solve the equation.

Neglecting the nonlinear "hole-burning" term in the denominator, this equation becomes linear, and matrix $T_{nn'}^\mu$ becomes diagonal. As a result, the "non-interacting" lasing thresholds and "non-interacting" 
lasing frequencies are obtained by  
\begin{align}
	d_0\gamma(\omega)=\eta_n(\omega). 
\end{align} 
To be more precise, we define the effective lasing frequency and effective pumping rate for each TCF by 
\begin{align}
&\omega_n^\mathrm{eff}(\omega)=\omega_a -\gamma_\perp \frac{\Re[\eta_n(\omega)]}{\Im[\eta_n(\omega)]}, \\
&d_{0n}^\mathrm{eff}(\omega)=-\Im[\eta_n(\omega)]\left(1+\frac{(\omega-\omega_a)^2}{\gamma_\perp^2}\right). 
\end{align}
The condition of $\omega_n^\mathrm{eff}(\omega)=\omega$ determines the noninteracting lasing frequency $\omega_{\mu_0}$ and 
$d_{0n}^\mathrm{eff}(\omega)$ at this $\omega$ is the non-interacting lasing threshold $d_0^{\mu_0}$. 
Then, we have a discrete set of the solutions $\{ \omega_{\mu_0},d_0^{\mu_0}\}$ ($\mu_0=1,2,...$) with $d_0^1<d_0^2<...$.   
The lowest noninteracting threshold $d_0^1$ becomes the actual lasing threshold.

The higher lasing thresholds are determined as follows.   
Suppose that there are $N_\mathrm{las}$ lasing modes at a given pumping $d_0$, the self-consistent lasing frequencies $\omega_\mu$ and field coefficients $c_n^\mu$, both of which depend on $d_0$, are obtained by solving Eq. \eqref{Eq:NLE}. 
Using the solution, we introduce the interacting threshold matrix as 
\begin{align}
&\tilde{T}_{nn'}(\omega)\nonumber\\
&\hskip5pt =\frac{\gamma(\omega)}{\eta_n(\omega)}
\int 	\dd\vb*{x}\theta(\vb*{x}) \frac{ \vb*{E}_{\eta_n}(\vb*{x};\omega)\cdot\vb*{E}_{\eta_{n'}}(\vb*{x};\omega)}{1+\sum_\nu |\gamma(\omega_\nu)\sum_{m}c_m^\nu \vb*{E}_{\eta_m}(\vb*{x};\omega_\nu)|^2}. 
\end{align}
By definition, one of the eigenvalues $\lambda_\sigma(\omega)$ satisfies 
\begin{align}
d_0\lambda_\sigma(\omega_\nu)=1 \quad (\nu=1,2,...,N_\mathrm{las}).
\end{align}
Other eigenvalues becomes real at particular frequencies $\omega_\mu'$ near the non-interacting lasing frequencies $\omega_{\mu_0}$  
other than  those of the $N_\mathrm{las}$ lasing modes.  
However, they satisfy $d_0\lambda_\sigma(\omega_\mu')< 1$. 
Increasing $d_0$ results in $d_0\lambda_\sigma(\omega_\mu')=1$ at a certain $d_0$. This $d_0$ is nothing but the $(N_\mathrm{Las}+1)$-th interacting threshold above which  $N_\mathrm{Las}+1$ lasing modes are obtained. 
In this way, increasing the pumping generally results in multimodal lasing.

The lasing field is expressed as a superposition of the TCF states. Therefore, the angular distribution of the lasing pattern is given by 
\begin{align}
 	\vb*{f}^\mu(\Omega)=\sum_n c_n^\mu \vb*{f}_{\eta_n}(\Omega;\omega_\mu). \label{Eq:angle}
\end{align}
The total radiation flux $F^\mu$ of the $\mu$-th lasing mode is expressed as
\begin{align}
 	F^\mu = \int \dd\Omega |\vb*{f}^\mu(\Omega)|^2. \label{Eq:flux}
\end{align}

\section{Spherical particle systems as a lasing platform}
Let us consider a cluster of non-overlapping dielectric spheres as a lasing platform. The spheres include a gain medium, and they are uniformly pumped.  The lasing comes from an optical resonance in the cluster via the multiple scattering among the spheres. To analyze the resonance, we first consider the system without the gain medium.   
The resonance in such a system can be evaluated from the optical density of states (DOS) $\rho(\omega)$ given by \cite{Ohtaka1982,Moroz1995}
\begin{align}
&\rho(\omega)=\frac{1}{\pi}\pdv{}{\omega}\mathrm{Im}[\log\det(t)-\log\det(1-tG)],\\
&t_{(\alpha\beta L)(\alpha'\beta' L')}=t_{\alpha l}^\beta\delta_{\alpha\alpha'}\delta_{LL'}\delta_{\beta\beta'},\\
&t_{\alpha l}^\mathrm{M}=
-\frac{\rho_\alpha j_l(\rho_b)j_l'(\rho_\alpha)-\rho_bj_l'(\rho_b)j_l(\rho_\alpha)}
{\rho_\alpha h_l(\rho_b)j_l'(\rho_\alpha)-\rho_bh_l'(\rho_b)j_l(\rho_\alpha)},\\
&t_{\alpha l}^\mathrm{N}=
-\frac{\frac{\rho_b}{\rho_\alpha} j_l(\rho_b)(\rho_\alpha j_l(\rho_\alpha))'
		-\frac{\rho_\alpha}{\rho_b}(\rho_bj_l(\rho_b))'j_l(\rho_\alpha)}
{\frac{\rho_b}{\rho_\alpha} h_l(\rho_b)(\rho_\alpha j_l(\rho_\alpha))'
		-\frac{\rho_\alpha}{\rho_b}(\rho_bh_l(\rho_b))'j_l(\rho_\alpha)},\\
&\rho_\alpha=q_\alpha r_\alpha, \quad \rho_b=q_b r_\alpha, \quad q_\alpha=\frac{\omega}{c}\sqrt{\epsilon_\alpha}, \\
&G_{(\alpha\beta L)(\alpha'\beta' L')}=G_{LL'}^{\beta\beta'}(\vb*{x}_{\alpha}-\vb*{x}_{\alpha'})(1-\delta_{\alpha\alpha'}),\\
&G_{LL'}^{\beta\beta'}(\vb*{x})=\frac{1}{l(l+1)}\sum_{L_1L_2}[(\vb*{P}^\beta)^\dagger]_{LL_1}G_{L_1L_2}(\vb*{x}) [\vb*{P}^{\beta'}]_{L_2L'},\\
&G_{LL'}(\vb*{x})=\sum_{L''}4\pi \ii^{l-l'+l''}h_{l''}(q_br) Y_{L''}^*(\Omega)\langle L|L''|L'\rangle,\\
&\langle L|L''|L' \rangle=\int\dd\Omega Y_L^*(\Omega)Y_{L''}(\Omega)Y_{L'}(\Omega),
\end{align}
where $\alpha$ is the index to specify the spheres, $L=(l,m)$ is the angular momentum index with $|m|\le l$, $\beta(=M,N)$ is the polarization index of the vector spherical wave, $j_l$ is the spherical Bessel function, $h_l$ is the spherical Hankel function of the first kind,   $Y_L$ is the spherical harmonics,  
and $\vb*{P}^\beta$ is the vector spherical wave expansion coefficient \cite{Ohtaka1979}.   The $\alpha$-th sphere has the dielectric constant $\epsilon_\alpha$, center position $\vb*{x}_\alpha$, and radius $r_\alpha$.

In the case of an isolated sphere, the DOS is written as 
\begin{align}
	&\rho(\omega)=\sum_{l\beta} (2l+1)\rho_l^\beta(\omega), \\
	&\rho_l^\beta(\omega)= \frac{1}{2\pi}\pdv{\mathrm{arg}(1+2t_{\alpha l}^\beta)}{\omega}. 
\end{align}
The Mie resonance of the isolated sphere results in   
a Lorentzian peak in the DOS spectrum. In the cluster, 
a series of Lorentzian peaks emerge from the multiple scattering among the spheres, and very high quality factors not available in the isolated sphere around the same frequency range can be obtained.

The resonances in the cluster can also be evaluated from the QNMs, which are the zeros of $\mathrm{det}(1-tG)$ in the complex plane of frequency.  
The real and imaginary parts of the complex eigenfrequencies of the QNMs determine the resonance frequencies and widths, provided that the eigenfrequencies are close to the real axis.

The TCF states are also given  by the zeros of $\mathrm{det}(1-tG)$ 
 in which the sphere dielectric constant $\epsilon_\alpha$ is replace by $\epsilon_\alpha+\eta$ and frequency is taken to be real-valued. Here, we assume that the spheres are uniformly pumped.  
Thus, the secular equation for the TCF states is given by 
\begin{align}
	&\sum_{\alpha'\beta' L'}(1-tG)_{(\alpha\beta L)(\alpha'\beta' L')}\psi_{\alpha'L'}^{\beta'}=0.  
\end{align}
This equation has a discrete set of solutions in the complex plane of $\eta$.  
The electric field of the TCF state is expressed as 	
\begin{align}
&\vb*{E}_\eta(\vb*{x};\omega)=\left\{
\begin{aligned}
	&\sum_{\alpha L}h_{l}(q_b|\vb*{x}-\vb*{x}_\alpha| )Y_{L}(\Omega(\vb*{x}-\vb*{x}_\alpha))\vb*{V}_{\alpha L}^\mathrm{out} \\
	& \hskip50pt \mathrm{outside\; the \; spheres}\\ 
	&\sum_{L}j_{l}(\tilde{q}_\alpha|\vb*{x}-\vb*{x}_\alpha|) Y_{L}(\Omega(\vb*{x}-\vb*{x}_\alpha))\vb*{V}_{\alpha L}^\mathrm{in} \\
	& \hskip50pt \mathrm{inside\; the \; \alpha-th \; spheres}
\end{aligned} 
\right.,\\
&\vb*{V}_{\alpha L}^\mathrm{out}=\sum_{L'\beta}[\vb*{P}^\beta]_{LL'} \psi_{\alpha L'}^\beta, \quad \vb*{V}_{\alpha L}^\mathrm{in}=\sum_{L'\beta}[\vb*{P}^\beta]_{LL'} s_{\alpha l'}^\beta \psi_{\alpha L'}^\beta, \\
&s_{\alpha l}^M= \frac{\ii/\rho_b}{\tilde{\rho}_\alpha j_l(\rho_b)j_l'(\tilde{\rho}_\alpha)-\rho_bj_l'(\rho_b)j_l(\tilde{\rho}_\alpha)},\\
&s_{\alpha l}^N= \frac{\ii/\rho_b}{\frac{\rho_b}{\tilde{\rho}_\alpha} j_l(\rho_b)(\tilde{\rho}_\alpha j_l(\tilde{\rho}_\alpha))'
	-\frac{\tilde{\rho}_\alpha}{\rho_b}(\rho_bj_l(\rho_b))'j_l(\tilde{\rho}_\alpha)},\\
&\tilde{\rho}_\alpha = \tilde{q}_\alpha r_\alpha,\quad  \tilde{q}_\alpha = \frac{\omega}{c}\sqrt{\epsilon_\alpha + \eta}.
\end{align}

At far fields, the angular distribution of the TCF state is expressed as 
\begin{align}
\vb*{f}_\eta(\Omega;\omega)=\frac{1}{q_b}\sum_{\alpha L}
(-\ii)^{l+1}Y_L(\Omega)\ee^{-\ii  q_b\hat{r}\cdot\vb*{x}_\alpha} \vb*{V}_{\alpha L}^\mathrm{out} 	
 \end{align}
Angular distribution and the lasing flux are given by Eqs. \eqref{Eq:angle} and \eqref{Eq:flux}, respectively.

\section{finite 1D lattice of spheres}

A typical example of BICs found in spherical systems is an infinite  1D lattice of spheres \cite{Bulgakov2017a,Bulgakov2018a}.  At the $\Gamma$ point, the symmetry-protected BIC emerges for specific modes of azimuthal angular momentum $m=0$. Off-$\Gamma$ BICs are also available.  
We thus consider a finite 1D lattice of $N$ spheres.

The reason why the BIC occurs for the Bloch modes of $m=0$ at the $\Gamma$ point is a symmetry mismatch between the Bloch modes and external radiation.  Thus, it is not limited to the 1D lattice of spheres, but can happen in other 1D structures such as distributed Bragg-reflector pillars \cite{Ochiai2018a,Sadrieva2019a}.

\begin{figure}
	\centering 
	\includegraphics[width=0.5\textwidth]{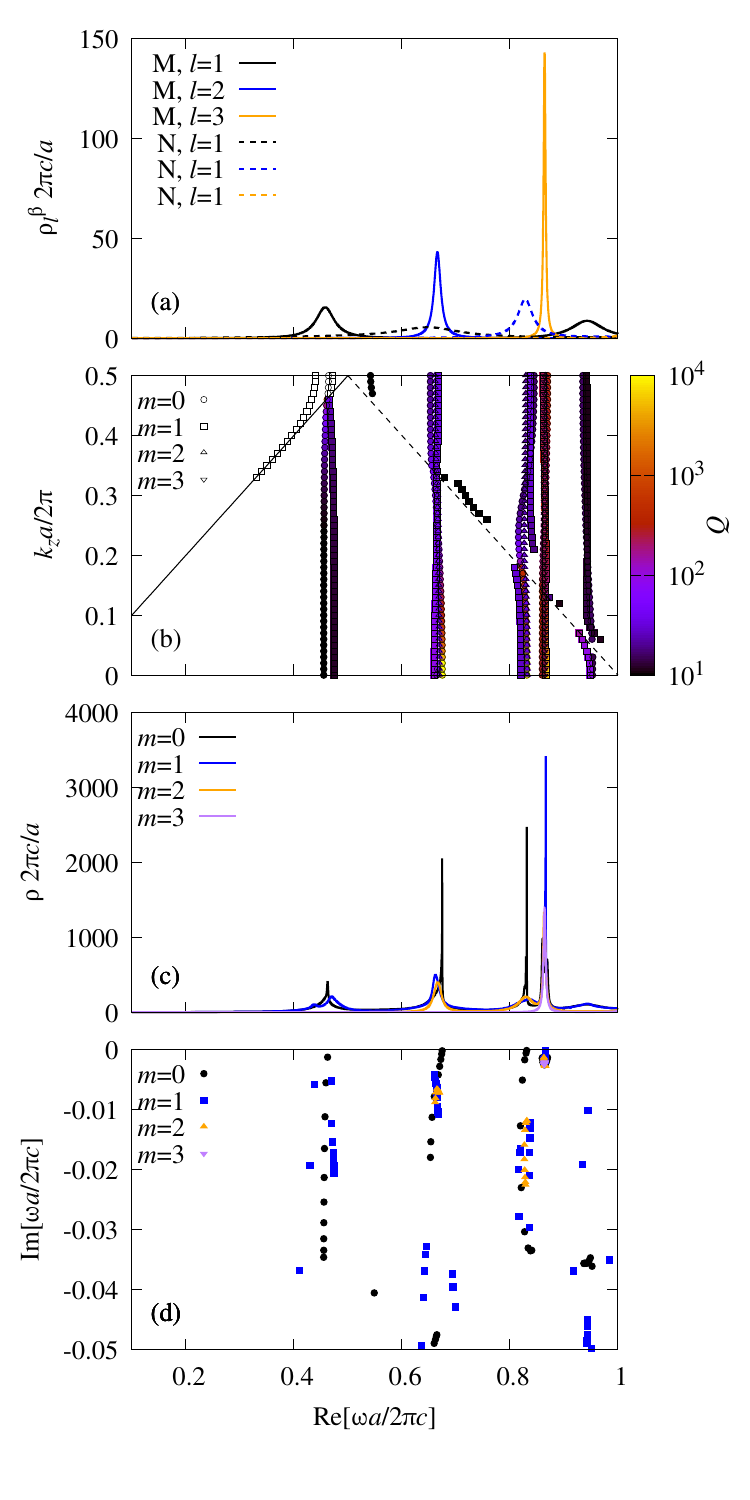}
	\caption{(a) Optical density of states $\rho_l^\beta(\omega)$ in an isolated dielectric sphere. (b) Photonic band structure of the infinite 1D lattice of the dielectric spheres. The quality factor of the photonic band modes are also shown in color. Solid and dashed lines represent the light line and the line of Bragg diffraction threshold, respectively.   (c) Optical density of states $\rho(\omega)$ in the finite 1D lattice of 10 dielectric spheres. (d) Complex eigenfrequencies of the quasi-normal modes in the finite 1D lattice of 10 dielectric  spheres.  The spheres are identical and have a dielectric constant of 12 and a radius $0.3a$, and are aligned periodically with a period of $a$ in the $z$ direction. The spectra are classified according to orbital angular momentum ($l$) and polarization in (a) and to azimuthal angular momentum ($m$) in (b,c,d). }
	\label{fig:ea12r03n10}
\end{figure}

Figure \ref{fig:ea12r03n10} shows the photonic band structure of the infinite 1D lattice of identical spheres in comparison to the optical DOSs of the isolated sphere and a finite 1D lattice of the spheres,  and QNMs of the finite 1D lattice of the spheres.  
The band structure, DOS, and QNMs are classified according to the azimuthal angular momentum $m$. The modes of $m=0$ are nondegenerate, whereas the modes of nonzero $m$ are doubly degenerate with those of $-m$. 
Since the photonic band modes are basically the tight-binding coupling among the Mie resonances of each sphere, 
the photonic bands are very flat and their eigenfrequencies are very close to those of the Mie resonances characterized by orbital angular momentum $l$ and polarization.  
The resonances become sharper in the finite 1D lattice.  
The sharpest peaks of the DOS correspond to the photonic band edges. 
In particular, the $m=0$ modes at the $\Gamma$ point can be the symmetry-protected BICs. 
We should note that off-$\Gamma$ BICs are not observed in the current parameters, consistent with the results in Ref. \cite{Bulgakov2017a}.

Since the sharp DOS corresponds to high $Q$, we consider the lasing by the high $Q$ modes. 
We then introduce the gain medium uniformly pumped in the spheres, whose spectral width covers the DOS peak around $\omega a/2\pi c=0.665$. 
by putting $\omega_a a/2\pi c=0.665$ and $\gamma_\perp a/2\pi c=0.1$.

\begin{figure}
	\centering
	\includegraphics[width=0.5\textwidth]{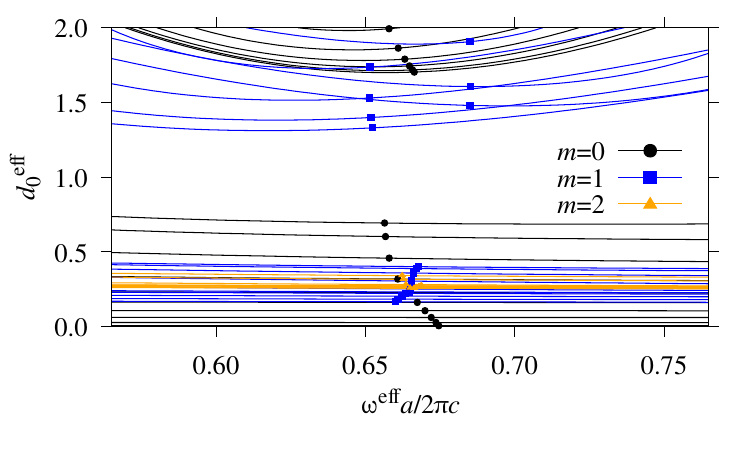}
	\caption{Dispersion of the TCF states concerning the effective lasing frequency and effective pumping rate. The non-interacting lasing thresholds are marked as the dots. The central frequency $\omega_a$ and transverse relaxation rate $\gamma_\perp$ of the gain medium are taken to be $\omega_aa/2\pi c=0.665$ and $\gamma_\perp a/2\pi c=0.1$. The same parameters as in Fig. \ref{fig:ea12r03n10} are taken. }
	\label{fig:ea12r03n10tcf}
\end{figure}

Figure \ref{fig:ea12r03n10tcf} shows the dispersion of the TCF states regarding the effective lasing frequency and effective pumping rate, by changing the frequency within the frequency window of the gain medium. 
The TCF states  shown in Fig. \ref{fig:ea12r03n10tcf} have $|\Re[\eta]|<5.5$, and other TCF with $|\Re[\eta]|>5.5$ are in the region outside the Fig.\ref{fig:ea12r03n10tcf}. Therefore, we can drop such TCFs in the expansion of Eq. \eqref{Eq:expandTCF}. 
The non-interacting lasing thresholds are also plotted as dots. The lowest one is the true lasing threshold.  
These points become lower in $d_0$ and denser with increasing system size. We thus might expect that the second threshold takes place just above the first threshold. However, it is not the case owing to the strong light confinement and nonlinear interactions by the hole-burning term.

\begin{figure}
	\centering
	\includegraphics[width=0.5\textwidth]{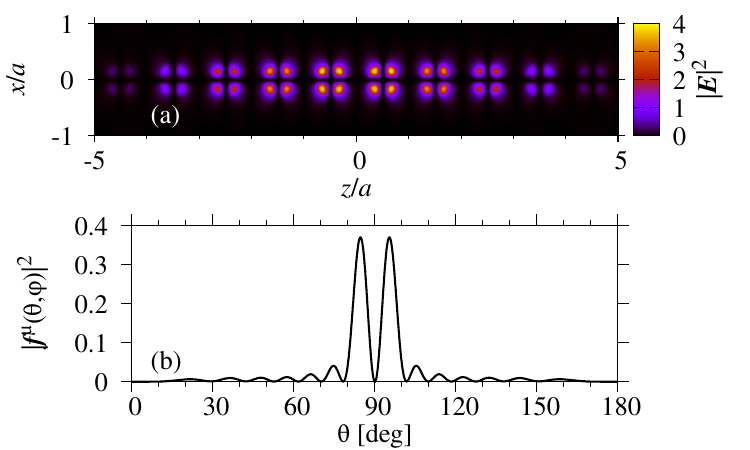}
	\caption{(a) Near-field pattern on the $zx$ plane ($y=0$) and (b) far-field angular distribution of the lowest-threshold lasing mode at $\omega a/2\pi c=0.6746$ and $d_0=0.0715$. The radiation field is normalized by $\sum_{\alpha L\beta}|\psi_{\alpha L}^\beta|^2=1$.  There is no azimuthal angle ($\phi$) dependence.  }
	\label{fig:ea12r03n10th1st}
\end{figure}

Figure \ref{fig:ea12r03n10th1st} shows the near and far-field patterns of the lowest threshold lasing mode.  
Since the relevant Mie resonance of the isolated sphere is of $l=2$ and the $M$ polarization, the near-field pattern reproduces the nodal structure of the $l=2$ Mie resonance.  The angular distribution in the far field exhibits the double peak around $\theta=90^\circ$ with the node there.  
This is a typical behavior of the lasing induced by a symmetry-protected BIC. The polar angle of $\theta=90^\circ$ corresponds to the $\Gamma$ point, where $Q$ is infinite and the coupling to the external radiation is forbidden. However, off the $\Gamma$ point, the $Q$ factor remains  very high, allowing for the coupling. That is why there is the double peak pushed by the suppression at the $\Gamma$ point.

\begin{figure}
	\centering
	\includegraphics[width=0.5\textwidth]{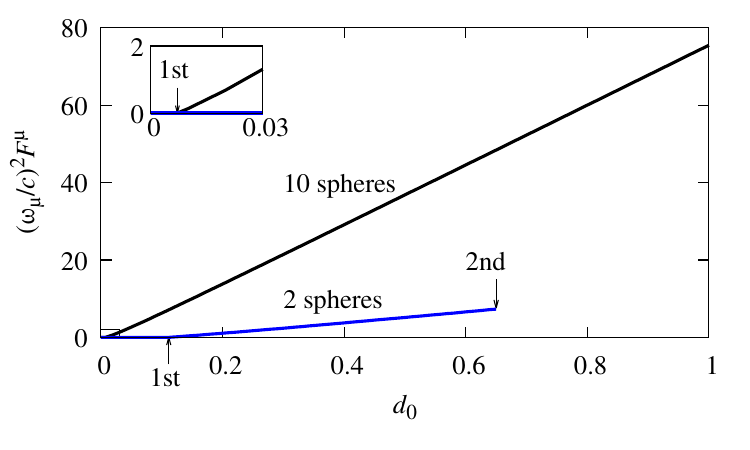}
	\caption{The lasing flux $F^\mu$ in the linear chain of 10 dielectric spheres. The lowest (first) threshold is at $d_0=0.00715$. For comparison, the lasing flux in the bisphere system with the same parameters is also shown. Although the bisphere exhibits the transition to the multimode regime, the single mode lasing is stabilized in the 10-sphere system.    
	}
	\label{fig:ea12r03n10salt}
\end{figure}

Figure \ref{fig:ea12r03n10salt} shows the lasing flux as a function of the pumping. 
The lasing flux behaves almost linearly with the pumping rate, exhibiting a  very low threshold.   The lasing frequency changes only $0.001\%$  from the non-interacting value $\omega^\mu a/2\pi c=0.6746$.  
The single-mode lasing is kept in the entire $d_0$ region above the threshold in the figure. 
Remarkably, there are 30 noninteracting thresholds below $d_0=1$. 
For comparison, in the bisphere case, the second threshold emerges at about $d_0=0.65$, showing the onset of the multimode lasing.

The final lasing pattern of the single-mode region does not change significantly from that in Fig.3.

\begin{figure}
	\centering
	\includegraphics[width=0.5\textwidth]{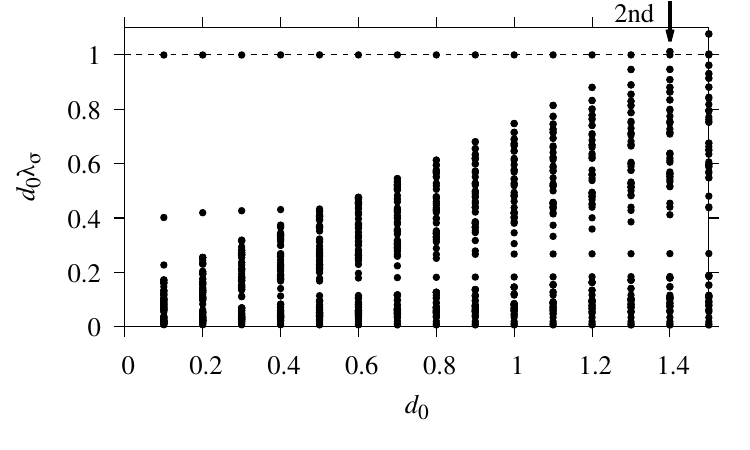}
	\caption{Real eigenvalues $\lambda_\sigma$ of the interacting threshold matrix in the finite 1D lattice of 10 spheres. The single-mode lasing is assumed in the evaluation of the interacting threshold matrix. The arrow indicates the second threshold.  }
	\label{fig:ea12r03n10saltintth}
\end{figure}

The real eigenvalues of the interacting threshold matrix assuming the single-mode lasing is shown in Fig. \ref{fig:ea12r03n10saltintth}
The eigenvalues at $d_0\lambda_\sigma=1$ are the lasing modes of interest. 
Those in $d_0\lambda_\sigma<1$ are unphysical modes, but can develop as the higher threshold lasing modes when they reach  $d_0\lambda_\sigma=1$.    
Even at $d_0=1$, the eigenmodes are far from the threshold value $d_0\lambda_\sigma=1$, and the second threshold does not take place. As can be seen in Fig. \ref{fig:ea12r03n25tcf}, there are many noninteracting thresholds below $d_0=1$.  
Therefore, the saturation gain by the hole-burning term is extreme in our systems. This behavior is consistent with the strong light confinement by the BIC.  
The second threshold emerges at about $d_0=1.4$. The second  threshold mode has the double degeneracy and its lasing frequency at the threshold is $\omega_{\mu=2}a/2\pi c=0.6606$, which is away from the lowest-threshold lasing frequency $\omega_{\mu=1}a/2\pi c=0.6746 $ at the second threshold. 
The double degeneracy and the close spacing both violate 
the basic assumption of the SALT, so that we stop 
analyzing further the multimodal region.

In this way, the symmetry-protected BIC results in three remarkable properties: 1. extremely low lasing threshold. 2. Double peaks with a node in the normal direction in the angular distribution. 3. extreme discrepancy between  non-interacting and interacting thresholds.

\section{finite 2D lattice of spheres} 
Next, we consider the square lattice of dielectric spheres. 
In this case, non-degenerate Bloch modes at the $\Gamma$ point are the symmetry-protected BICs. Off-$\Gamma$ BICs are also available \cite{Kostyukov2022,Ochiai2024}. 

\begin{figure}[h!]
	\centering
	\includegraphics[width=0.5\textwidth]{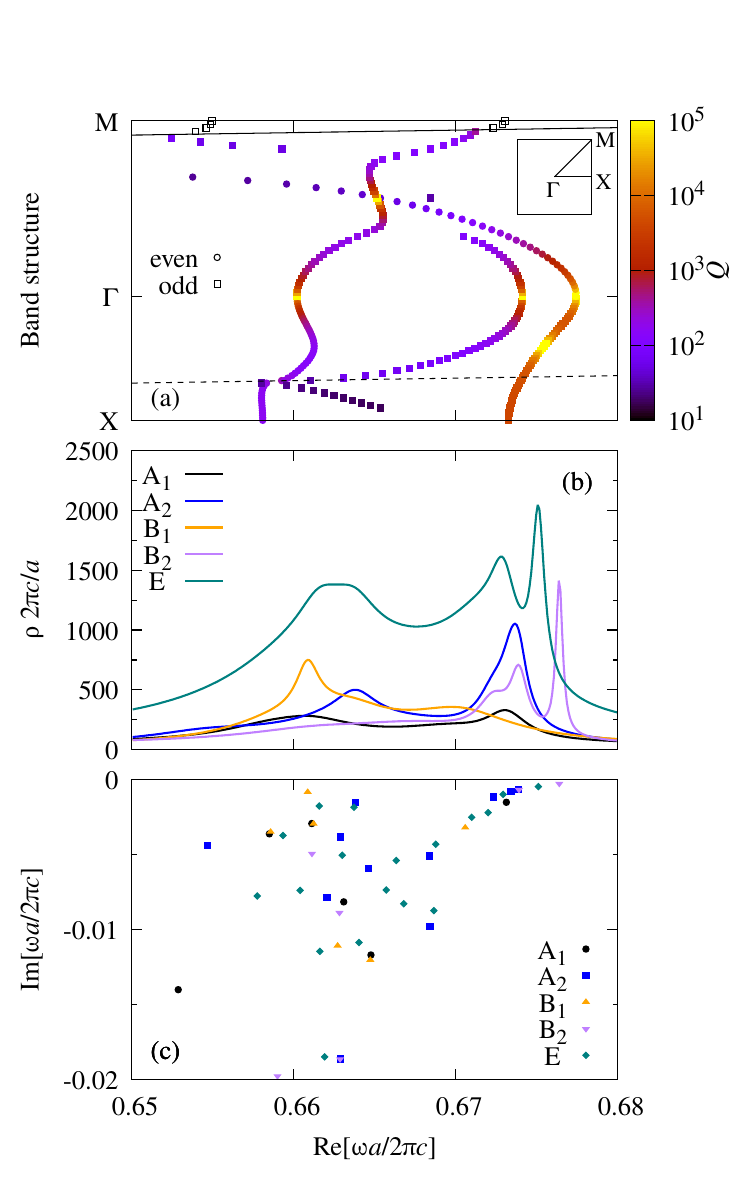}
	\caption{(a) Photonic band structure in the square lattice monolayer of dielectric spheres. The eigenmodes are classified according to the in-plane parity relevant to the $\Gamma$X and $\Gamma $M directions. The $Q$ factor is also plotted in the color scale. The solid and dashed lines represent the light line and the Bragg diffraction threshold line, respectively. The inset shows the first Brillouin zone. (b) The optical DOS $\rho(\omega)$ in the finite square lattice made of 25($=5\times 5$) spheres. The DOS is divided into the irreducible representations of $C_{4v}$. The two sharp peaks of the $B_2$ and $E$ representations correspond to the at-$\Gamma$ and off-$\Gamma$ BICs. (c) QNM eigenfrequencies in the finite square lattice. The QNMs  are also classified by the irreducible representations of $C_{4v}$.  The sphere's dielectric constant and radius are 12 and $0.3a$, where $a$ is the lattice constant of the square lattice. The parity-odd modes concerning the plane of the monolayer are considered. 
	\label{fig:sqbd}}
\end{figure}

Figure \ref{fig:sqbd} shows the photonic band structure of the square lattice of dielectric spheres in comparison to the optical DOS and QNMs in the finite square-lattice system.   
The band structure exhibits a series of symmetry-protected BICs at the $\Gamma$ point
as the bright spots in the band diagram.    
Off-$\Gamma$ BICs are also found along the $\Gamma$X and $\Gamma$M directions.  
The optical DOS exhibits two marked peaks for $B_2$ and $E$ modes, where the eigenfrequencies of the QNMs of the exact representations become closer to the real axis in the complex frequency plane. These peaks correspond to the two BIC points in the band structure. 
Since the DOS peak frequency of the $E$ mode is not at the band edge, the peak is not simply due to the off-$\Gamma$ BIC in $\Gamma$X, but includes other Bloch modes in the first Brillouin zone.

\begin{figure}
	\centering
	\includegraphics[width=0.5\textwidth]{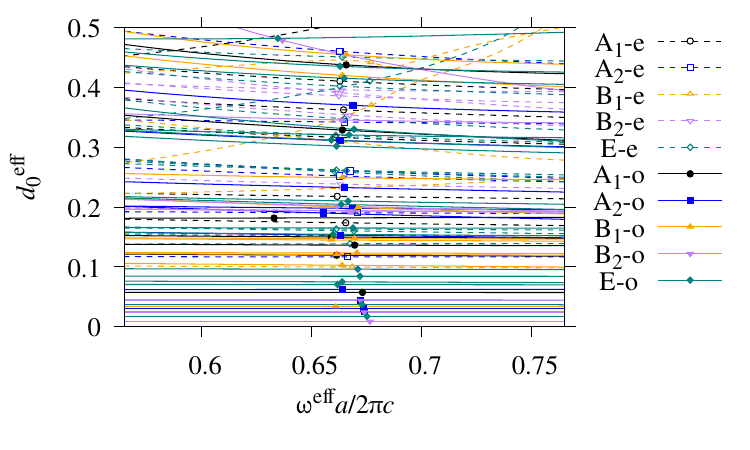}
	\caption{Dispersion of the TCF states regarding the effective lasing frequency and pumping rate in the finite square lattice made of 25(=5$\times$ 5) spheres. The non-interacting lasing thresholds are denoted as the dots. The TCF states are classified according to the irreducible representations of $C_{4v}$ together with the parity concerning the plane of the monolayer. For instance, symbol  $A_1\mathrm{-e}$ represents the parity-even mode of the $A_1$ representation of $C_{4v}$.  
	\label{fig:ea12r03n25tcf}}
\end{figure}

Next, we consider the TCF states and resulting non-interacting thresholds. 
Figure \ref{fig:ea12r03n25tcf} shows the TCF map regarding the effective lasing frequency and pumping rate. 
As suggested in the DOS and QNM spectra of Fig. \ref{fig:sqbd}, the lowest threshold is at $d_0= 0.0092$ of the $B_2$ odd mode. The second lowest one is at $d_0=0.017$ of the $E$ odd mode. There are many low-threshold modes among various irreducible representations.

\begin{figure}
	\centering
	\includegraphics[width=0.5\textwidth]{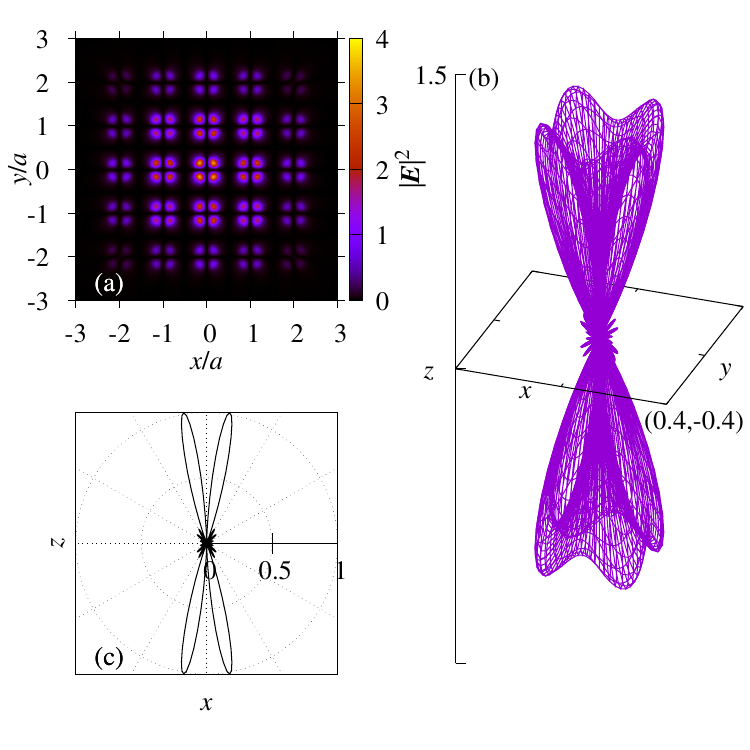}
	\caption{(a) Near-field configuration of the electric field of the lowest-threshold lasing mode in the finite (5$\times$5) square lattice of the dielectric spheres. The plane of $z=0$ is assumed. (b) Far-field angular distribution of the lowest-threshold lasing mode. (c) The far-field angular distribution in the $xz$ plane. The lasing mode is normalized as 
	$\sum_{\alpha L\beta}|\psi_{\alpha L}^\beta|^2=1$. In (b) and (c), the angular distribution is given by putting  $r=(\omega_\mu/c)^2 |\vb*{f}^\mu(\Omega)|^2$ in the polar coordinate. }
	\label{fig:ea12r03n25th1st}
\end{figure}

The near- and far-field patterns of the lowest-threshold lasing mode are shown in Fig. \ref{fig:ea12r03n25th1st}
The resonance feature inside the spheres is very close to that in Fig. 3(a). The angular distribution of the lasing has a ring-like peak around the normal direction to the monolayer with a node in the normal direction. This feature also resembles that in Fig. 3(b). Both the threshold lasing modes are the symmetry-protected BICs at the $\Gamma$ point in the infinite system-size limit. Therefore, the emission in the normal direction is prohibited. 
If we increase the system size, the angular distribution of the lowest-threshold lasing mode becomes sharper toward the normal direction, but still has a node there. In this way, the resulting laser is expected to be more surface emitting as the system size increases.

\begin{figure}
	\centering
	\includegraphics[width=0.5\textwidth]{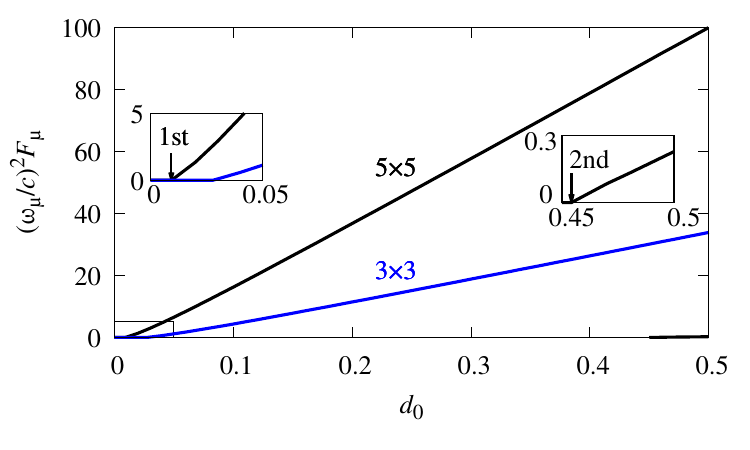}
	\caption{Lasing flux $F^\mu$ as a function of pumping $d_0$. For comparison, the lasing flux in $3\times 3$ lattice of the spheres is also shown. The Insets show the enlarged graphs around the thresholds.  
     The lasing frequency $\omega_\mu$ also depends on $d_0$. However, it changes less than 0.001$\%$ from the threshold value $\omega_\mu a/2\pi c=0.6763$ in the $d_0$ range shown.  
	 }
	\label{fig:ea12r03n25salt}
\end{figure}

The lasing flux is shown in Fig. \ref{fig:ea12r03n25salt} 
While there are so many noninteracting thresholds below $d_0=0.5$, the single-mode lasing sustains up to $d_0=0.454$. 
Comparing with the lasing flux in the  $3\times 3$ lattice of the spheres, the slope efficiency {\it per sphere} does not change so much. This trend is consistent with the fact that the radiation field of the BIC is confined only vertically. The horizontal confinement is not absent in the BIC.   
Therefore, the modal volume of the BIC is much larger than that of typical photonic-crystal defect mode cavities. 
However, this property is crucial for making the BIC laser spatial-dispersion free or surface emitting.

\begin{figure}
	\centering
	\includegraphics[width=0.5\textwidth]{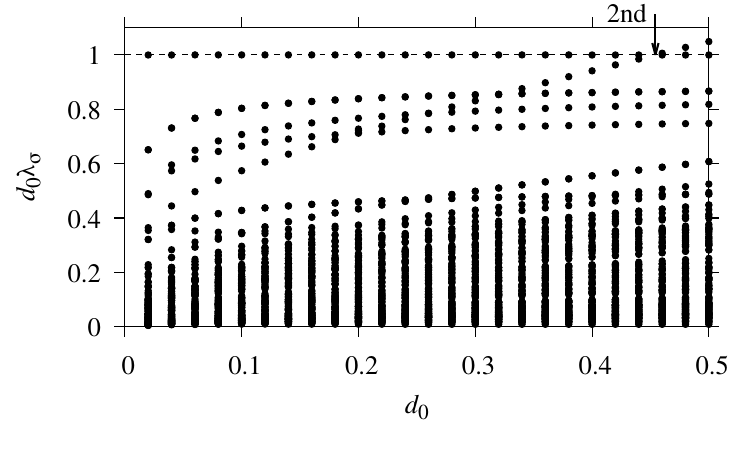}
	\caption{Real eigenvalues $\lambda_\sigma$ of the interacting threshold matrix as a function of pumping $d_0$, assuming the single-mode lasing in the evaluation of the interacting threshold matrix. The second threshold emerges around $d_0=0.454$.  }
	\label{fig:ea12r03n25saltintth}
\end{figure}

The real eigenvalues of the interacting threshold matrix is shown in  Fig. \ref{fig:ea12r03n25saltintth}
In this case, the second threshold emerges around $d_0=0.454$. 
The second lasing mode is doubly degenerate, and the lasing frequency  at $\omega a/2\pi c=0.6608$ at the  threshold. 
Therefore, it is relevant to the off-$\Gamma$ BIC mode. 
However, the SALT does not handle the closely spaced multimode lasing frequencies; we have to stop the present analysis.

\section{Summary and discussions}

In summary, we have investigated the BIC lasing in the finite lattices of the dielectric spheres, using the SALT. The BICs of specific Bloch modes in the corresponding infinite lattices turn into a dense distribution of high $Q$ QNMs around the BIC frequencies. The highest $Q$ mode corresponds to the BIC and acts as the lasing mode with a very low threshold.  In contrast to the closely spaced noninteracting thresholds, the single-mode lasing is stabilized far beyond the higher  noninteracting thresholds, owing to a strong nonlinearity of the hole-burning term.

There are many issues that remain unsolved. The most important one is a possible stabilization criterion. 
In the 1D system, the stabilization of the 10-sphere system is much stronger than that of the bisphere system.  In contrast, in the 2D system, the stabilization of the 5$\times$5 system is weaker than in the 3$\times$3 system. In the infinite periodic system, it is shown that the stabilization of the band-edge lasing takes place if the gain-medium frequency $\omega_a$ is in the band gap \cite{Benzaouia2020b}. In our case, $\omega_a$ is within a band. Therefore, there might be some critical system sizes above which the single mode tends to unstabilize. A detailed investigation of it is in order.  

Related to this issue, finite-size effects of off-$\Gamma$ BIC lasing are also important. In our parameter setting, the off-$\Gamma$ BIC does not emerge as the lowest threshold lasing mode. To realize the off-$\Gamma$ BIC, we generally require large system sizes. There, the band curvature is shown to 
be critical for the stability.

There is another stabilization scenario of the single-mode lasing, related to the non-Hermitian skin effect \cite{Zhu2022}. Although it is not related to the BICs, it is interesting to investigate the possible combined effects of the BICs and non-Hermitian skin effect. 

Another important issue is the practical treatment of the multimodal region within the SALT. In the structures under study, the second threshold emerges at a specific pumping rate. The second threshold mode is degenerate in the 1D lattice system and nondegenerate in the 2D lattice systems under the current parameter setting.  
The lasing frequency of the second mode is very close to the lowest one, so that the basic assumption of the SALT is violated.  Although a time-domain simulation of the Maxwell-Bloch equation is available, there should be a frequency-domain approach extending the SALT.

In our systems, we impose spatial symmetry to reduce the computational resources and symmetry assignment.  In practice, the symmetry is broken by structural imperfection or disorder. The off-$\Gamma$ BICs are generally topologically protected. Even for symmetry-protected BICs, those of higher vortex charges can split to the BICs of lower vortex charges \cite{Yoda2020}. Therefore, many BICs can be preserved after breaking the spatial symmetry. The BIC lasing after breaking the symmetry is thus available.  

Other issues such as linewidth, non-uniform pumping, 
laser chaos, etc, are to be investigated. 

We hope this paper stimulates further investigation of the BIC lasing toward the ultimate laser with zero threshold and zero linewidth.

\begin{acknowledgments}
This work was supported in part by JSPS KAKENHI grant number 22K03488 in the initial stage.   	
\end{acknowledgments}

%\bibliography{../../Database/library,../../Database/library_add}

\end{document}